# Transport properties and electrical device characteristics with the TiMeS computational platform: application in silicon nanowires

D. Sharma[1‡], L. Ansari[1‡], B. Feldman[2], M. Iakovidis[1], J. C. Greer[1] and G. Fagas[1*]

[1] Tyndall National Institute, University College Cork, Cork, Ireland

[2] Department of Materials and Interfaces, Weizmann Institute of Science, Rehovoth 76100, Israel

[*] Correspondence to: Georgios.Fagas@tyndall.ie

**Abstract**

Nanoelectronics requires the development of *a priori* technology evaluation for materials and device design that takes into account quantum physical effects and the explicit chemical nature at the atomic scale. Here, we present a cross-platform quantum transport computation tool. Using first-principles electronic structure, it allows for flexible and efficient calculations of materials transport properties and realistic device simulations to extract current-voltage and transfer characteristics. We apply this computational method to the calculation of the mean free path in silicon nanowires with dopant and surface oxygen impurities. The dependence of transport on basis set is established, with the optimized double zeta polarized basis giving a reasonable compromise between converged results and efficiency. The current-voltage characteristics of ultrascaled (3 nm length) nanowire-based transistors with p-i-p and p-n-p doping profiles are also investigated. It is found that charge self-consistency affects the device characteristics more significantly than the choice of the basis set. These devices yield source-drain tunneling currents in the range of 0.5 nA (p-n-p junction) to 2 nA (p-i-p junction), implying that junctioned transistor designs at these length scales would likely fail to keep carriers out of the channel in the off-state.

## 1. Introduction

The continuous miniaturization of electronic transistors pursued by semiconductor industries enables more functionality for a fixed die area. However, scaling devices to sub-deca nanometer gives rise to short-channel and quantum tunneling effects that degrade device performance[1,2] and to keep up with scaling research has motivated new transistors designs that use a range of nanostructured materials.[3] These types of devices require for their description taking explicitly into account new physical phenomena and materials properties at the nanoscale. For example, strong quantum confinement in one- and two- dimensions occurs respectively in ultrathin body and nanowire-based field-effect-transistors (FETs)[2,4] and the high surface to volume ratio allows for the manifestation of size effects [5,6] and volume inversion.[7,8] Direct source-drain quantum mechanical tunneling[3,9] and gate-tunneling leakage arise,[10] obstructing the way to reducing power consumption.[11,12] Manifestations of differing properties of nanoscale materials compared to their bulk counterparts have also been demonstrated. For the prototypical material of silicon nanowires, surface functionalization schemes result in tuning the electronic and transport properties,[13,14,15] the effective masses of charge carriers become heavier,[16] dopants may deactivate,[17] and the deformation potentials and electron-phonon scattering can become highly anisotropic.[18]

Various semi-classical methods have been elaborated to simulate the current-voltage of conventional transistors and are used to reduce costs and shorten the design cycle.[19] The need to develop such *a priori* technology evaluation that extends to the nanoscale is significant as the traditional trial-and-error experimental design of nanodevices becomes more time consuming and expensive. Over the last decade the description of electronic quantum transport based on computational methods has become one of the core topics in atomic-scale modeling and device simulations,[20] and the explicit electronic structure of materials has been considered from approximate methods that use empirical bulk parameterization[15] to first-principles approaches[21,22,23], and approximations thereof.[13,23,24,25]

In this paper, we first discuss our own quantum transport implementation (TiMeS – <u>T</u>ransport <u>i</u>n <u>Me</u>soscopic <u>S</u>ystems). Separating the step that provides the electronic structure description from the transport module, TiMeS offers a cross-platform solution that allows efficient calculations of materials transport properties and realistic device simulations to extract current-voltage and transfer characteristics. Previously, TiMeS was applied to predict transistor behavior in junctionless nanowire-based gate-all-around (GAA) architectures with just a 3 nm gate length.[9,25] The excellent device characteristics at such ultra-scaled dimensions have been recently confirmed.[26] The simulation results were obtained using an approximate Density-Functional Tight-Binding (DFTB) approach based on the Harris-Foulkes functional of density functional theory (DFT).[27] Here, we make use of the TiMeS modularity to interface with a DFT package based on numerical atomic orbitals. We demonstrate TiMeS flexibility by applying it to study the basis set dependence of the mean free path in silicon nanowires with dopant and surface oxygen impurities and the current-voltage characteristics of ultrascaled nanowire devices.

Theoretical studies based on first-principles DFT is one of the widely used methods to describe accurately the atomic geometry and to provide materials design guidelines in an affordable computational time without introducing system dependent parameters. In a recent study,[28] we benchmarked numerical atomics orbitals (NAOs) against plane waves (PWs) basis sets and discussed the impact of basis set on the electronic properties of SiNWs. PWs are simpler to converge, however computational requirements have limited their use in studies of quantum transport. On the other hand, basis sets made of atomic orbitals (AOs) can be more efficient motivating their implementation for large-scale order-N calculations.[29]

There have been numerous studies comparing the structural properties and electronic spectra resulting from PW and AO implementations (see e.g., Ref. 28 and references therein). Surprisingly, with few exceptions a similar elaboration on the impact for transport properties has attracted little attention despite the need to reach the same level of confidence for device design and evaluation.

Strange *et al.* performed benchmark calculations of the transmission spectrum for a set of five single-molecule junctions.[30] Using a Wannier transformation on DFT Hamiltonians in the PW basis sets enabled them to calculate the transmission function and to compare with a NAOs transport implementation. They concluded that a double zeta polarized basis sets suffices for the particular systems. This confirmed earlier work by Bauschlicher *et al.* on the gold-benzene-1,4-dithiol-gold junction.[31] In Ref. 32 Driscoll and Varga study the dependence of quantum conductance on basis sets by comparing localized basis sets with extended non-localized polarized basis. They find that convergence with localized basis sets is more demanding due to sensitivity in describing the self consistent potential. Hermann *et al.* also identify issues with the use of large non-orthogonal Gaussian-type AOs.[33] They show that basis sets of triple-zeta quality or higher can result in an artificially high transmission. These results imply that despite the extensive study over decades of the convergence properties of AOs in quantum chemistry[34], it is necessary to check the transferability of tabulated AOs as well as the construction of NAOs in different chemical environments for *a priori* evaluation of nanoscale devices.

The structure of the paper is as follows. In the next section the background on the computational methodology to analyze materials properties and device characteristics at the nanoscale is expanded. In Section 2.2 in particular, we provide the implementation details of the TiMeS cross-platform simulation tool. This is based on employing Hamiltonian matrix descriptions from different electronic structure codes and combining these with Green's functions methods to calculate the quantum-mechanical scattering matrix (S-matrix). Charge self-consistency in the presence of applied voltages is treated within the non-equilibrium Green's functions framework. In Section 3.1, TiMeS is applied to the calculation of the mean free path and the basis set convergence for silicon nanowires with dopant and surface oxygen impurities is established. Optimized double zeta polarized basis sets give a reasonable compromise between converged results and efficiency. The current-voltage characteristics of ultrascaled (3 nm long) nanowire-based transistors with p-i-p and p-n-p doping profiles is investigated in Section 3.2. The quantitative interplay between basis set dependence and charge self-consistency is analyzed and it is found that the latter has a more pronounced effect on the device characteristics. Interestingly, these devices yield relatively large source-drain tunneling in the off-state (currents in the range of of 0.5 nA to 2 nA for the p-n-p and p-i-p junctions, respectively), which can have a detrimental effect in the device performance. Section 4 concludes with a brief discussion of the results.

## 2. Computational Methodology

2.1 *Geometry and electronic structure*

The geometry and electronic structure are calculated using DFT. Throughout the simulations the PBE generalized gradient approximation (GGA) to the exchange correlation potential is used together with norm-conserving pseudopotentials. We focus on hydrogenated silicon nanowires grown along the [110] crystallographic direction with lattice constant of 3.84 Å along with the wire axis and diameter equal to 1.15 nm. Impurities are introduced in the supercell by either substitution of silicon atoms in the pristine structure with dopants (see Figure 1) or changing the surface termination and introducing oxygen defects for the model described in Ref. 14. The structures were geometry optimized with force threshold of 0.01

eV/´ and using optimized double zeta polarized basis set for both silicon and impurity atoms. Surface-passivating hydrogens are treated in a minimal basis set. The total energy using these bases are a good approximation to the complete basis set limit as shown previously.[28] The supercell consisted of eleven unit cells and its size extended by 42.24 Å x 25 Å x 25 Å to introduce a vacuum separation between periodic images of the nanowires. Monkhorst-Pack k-point sampling was applied on a 4x1x1 grid with the higher sampling aligned to the nanowire axis.

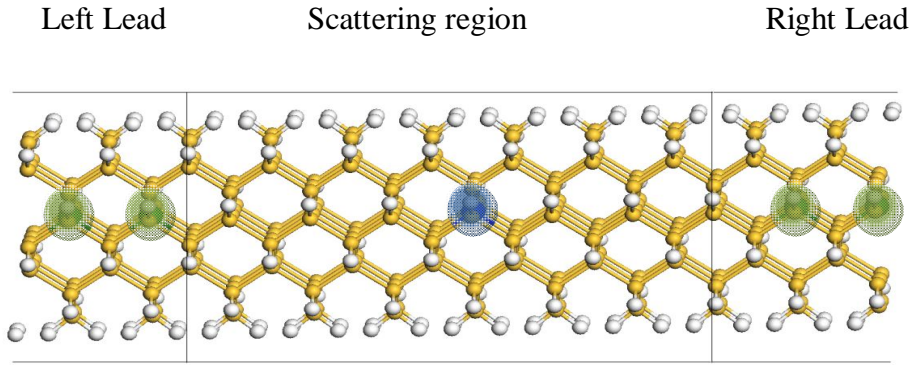

**Figure 1.** Schematic setup of one of the investigated nanowire structures. A silicon nanowire grown along the [110] direction is depicted with p-n-p doping profile. Substitutional boron (green sphere on line) is used to dope the semi-infinite periodic leads whereas the "scattering region" incorporates a phosphorus dopant (blue sphere on line).

Here, we employ the DFT implementation in the OpenMX code as developed by Ozaki.[35,36] This expresses the Kohn-Sham wavefunction $\psi_i$ as a linear combination of numerical atomic orbitals (NAOs):

$$\psi_i(\vec{r}) = \sum_n \sum_{\mu=1}^{K} C_{i\mu} \varphi_\mu(\vec{r} - \vec{r}_n) \quad (1)$$

The set $\{\varphi_\mu; \mu =1, 2…K\}$ of pseudoatomic wavefunctions, referred as primitives, is obtained by solving the Schrödinger equation for an atom in a slightly modified environment that accounts to some extent for the orbital relaxation when bonding; $n$ is the site index and $\mu$ includes the angular momentum *l*, magnetic quantum number *m* and the multiplicity index *p*, namely, $\mu \to p, l, m$. The basis set notation smpm'dm" is used to indicate that m, m' and m" functions are used to expand s, p, and d orbitals, respectively, relating to the quantum chemistry notation of single-zeta, double-zeta and so on, depending on the number of primitives per valence state. In practice the NAOs $\{\varphi_\mu\}$ are further expressed as a linear combination of pseudoatomic orbitals with the same *(l,m)* index pair, namely,

$$\varphi_\mu(\vec{r}) = \sum_q \alpha_{\mu q} \chi_\eta(\vec{r}) \qquad (2)$$

with $\eta \to q, l, m$ and $\mu \to p, l, m$. This leads to contracted or optimized numerical atomic orbitals, thereby improving the numerical efficiency and accuracy.[28,35] Orbital optimization reduces the size of the basis and reduces effects associated with basis sets overcompleteness. The notation $\varphi$nm indicates abbreviation of basis sets, e.g., s32 means that two optimized s orbitals are constructed from three primitive functions.

*2.2 Transport properties*

We have developed a modular transport simulator TiMeS that allows us to assess both the intrinsic electronic transport properties of materials and the electrical characteristics of devices. Our transport module requires a single particle Hamiltonian of the system expressed in a localized basis set. Based on Green's function techniques, the S-matrix and hence the transmission *T(E)* of charge carriers injected at energy *E* can be calculated from a single-particle Hamiltonian *H* in either the low-bias (non-self-consistent) approximation, or in fully self-consistent (SC) non-Equilibrium Green's function (NEGF) theory. The interfacing across various electronic structure platforms and the SC NEGF extensions build on previous work on the transport properties of mesoscopic systems[38,39] and molecular junctions[24,40].

*2.2.1 Low-bias regime*

We emphasize that the non-self-consistent (NSC) version of TiMeS operates as a post-processing step after an electronic structure computation and allows for extracting transport properties in the low-bias regime. Importantly, TiMeS is completely modular in that it requires, and has, no information about the representation used by the electronic structure code *except* for the single-particle (typically Kohn-Sham) Hamiltonian matrix $H_{ij}$ and overlap matrix $S_{ij} = <i|j>$ elements. TiMeS currently has interfaces to accept this information from OpenMX,[35-37] DFTB+,[27] and the Quantum Espresso plane-wave DFT code via transformation to Wannier orbitals.[41]

Like other localized-orbital electronic transport codes based on Landauer or NEGF theory,[21-24] TiMeS calculates the self-energy of the semi-infinite electrodes based on the surface Green's function for the given 'on-site' and 'hopping' Hamiltonian and overlap matrices for the electrodes.[43] The entire transport region is broken into three sub-regions: the two electrodes and a scattering region (see Figure 1). The electrodes are semi-infinite repetitions of periodic cells. These cells and the scattering region must be "principal layers" such that overlap elements vanish for orbitals separated beyond the adjacent principal layers. Therefore, just one 'hopping' Hamiltonian is needed for each electrode.

With the non-Hermitian self-energies $\Sigma_{L,R}$ to represent semi-infinite electrodes, TiMeS performs inversion to solve for the Green's function $G(E) = (E - H - \Sigma_L - \Sigma_R)^{-1}$. This step is performed independently for each energy E; hence, a parallel MPI implementation is used to reduce the computational cost. From G(E) the scattering matrix is calculated using the channel eigenstates of the leads[38,44]. This approach enables monitoring the computational flow and numerical stability by checking the unitarity of the S-matrix at each energy step. It also allows decomposition into individual channel contributions, thereby yielding direct information on the effect of the various scattering mechanisms between channels.[38,39,44] The explicit calculation of the channel eigenstates can be used further to obtain the band structure of the leads including the forbidden energy domain where modes are decaying exponentially along the periodic direction.[40] Additional information is obtained by analysis of the density of states[14] as extracted from diagonal matrix elements of the Green's functions for any section of the device.

### 2.2.2 Self-consistent transport using NEGF

Starting with this low-bias algorithm, we extended TiMeS to perform fully self-consistent NEGF calculations in a fashion retaining TiMeS' modularity. Following Ke *et al.*[23] for applying boundary conditions to the electrode regions allows us to implement NEGF independently of the electronic structure step. Solution for the Green's function G from $H_{KS}$ gives equivalent information to solving the KS equations, so G contains information on the state of the system.[23,44] TiMeS calculates the electronic density matrix (DM) from the (typically KS-DFT) Hamiltonian output by the electronic structure code. But note that unlike the electronic structure step, which is typically carried out with periodic or cluster boundary conditions (BCs), TiMeS finds the DM for the *open* system, with self-energies representing the semi-infinite electrodes.

The DM is calculated according to:

$$DM = -\frac{1}{\pi} \int_{-\infty}^{+\infty} dE \ \text{Im}[G_C^r(E) f^L(E-\mu_L;T_e)] + \frac{1}{2\pi} \int_{-\infty}^{+\infty} dE \ [G_C^r(E)\Gamma_R(E)G_C^r(E)^\dagger][f^R(E-\mu_R;T_e)-f^L(E-\mu_L;T_e)] \quad (3)$$

where $G_C^r$ is the retarded Green's function in the scattering region, f is the Fermi function at electronic temperature $T_e$ and $\mu_{L,R}$ are the chemical potentials of the two leads. The spectral density is defined via $\Gamma_{L,R} = i[\Sigma_{L,R} - \Sigma_{L,R}^\dagger]$. The first term in Equation (3) is the equilibrium

(linear-response) contribution to the DM, and the second term is the non-equilibrium (polarization and current flow) contribution, which vanishes when $\mu_L = \mu_R$. The integration is performed by Gaussian quadrature, and the equilibrium component may also use complex contour integration (starting appreciably below the lowest energy band), such that each integral usually requires evaluation of G(E) at ~50 or fewer energies E. Note that even in a zero-bias calculation, the linear-response density in NEGF can differ from the non-self-consistent result because of the inclusion of the self-energies $\Sigma_{L,R}$, representing open BCs rather than the periodic BCs of the electronic structure step. In the limit of an ideal calculation, the central region should be made long enough that the BCs are virtually irrelevant, but this is not always the case in practice. In the case of an applied voltage bias, convergence is ensured by extending the scattering region.

The charge density can be computed straightforwardly from the DM according to

$$\rho(r) = \sum_{\mu,\nu} \varphi_\mu^*(r) Re\big[(\text{DM})_{\mu,\nu}\big] \varphi_\nu(r) \quad (4)$$

and this density is used in the electronic structure code to extract the new device region Hamiltonian matrix. As a general principle, codes based on KS-DFT can find the entire system Hamiltonian $H_{KS}$ from knowledge of just the electronic density. In addition, the DFT-based tight-binding code DFTB$^+$ [27] can also be re-started using atomic Mulliken charges as input. The source code of the electronic structure packages is modified to read in the re-start density variables before proceeding with a single KS step to find the new $H_{KS}$. To facilitate calculations of gated devices, we also modified OpenMX to include the gating potential from a cylindrical GAA shell of charge.[45] Source-drain voltages are applied by including a potential step in the electrodes and then the charge density is iterated to self consistency.[23]

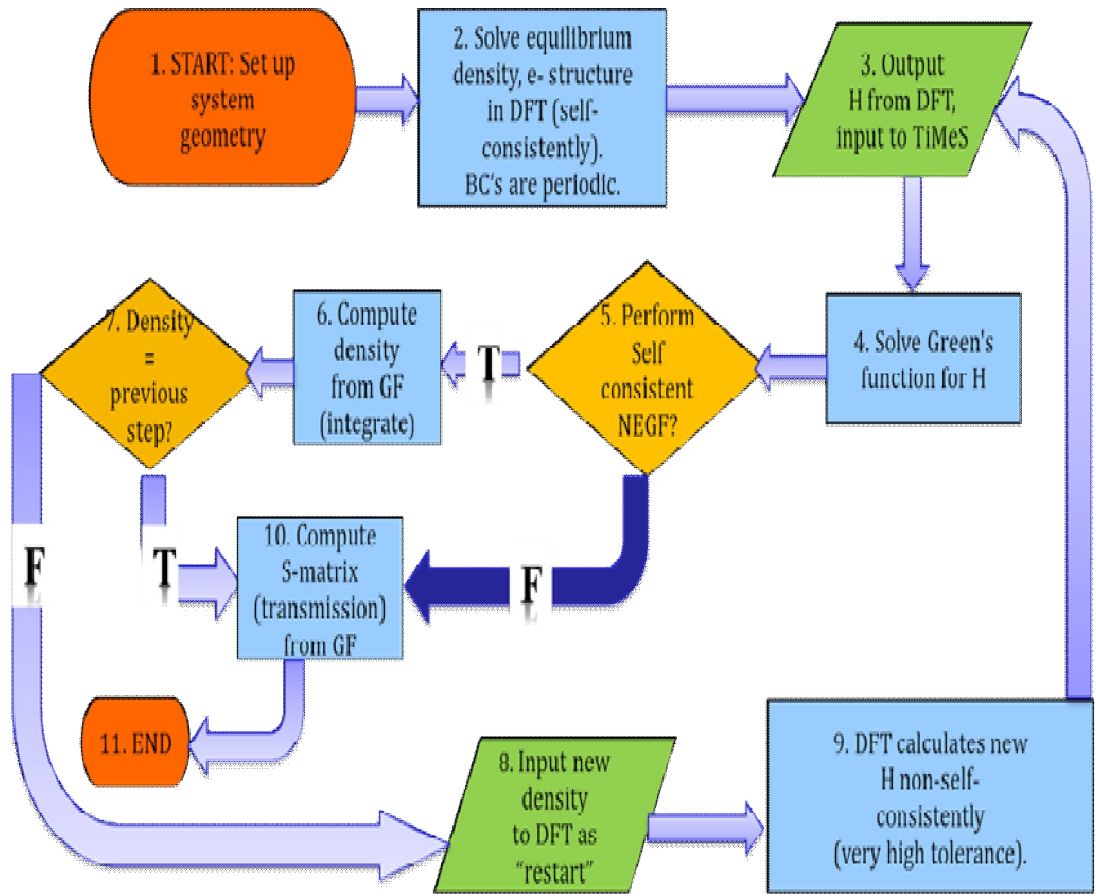

**Figure 2.** The TiMeS flowchart shows the cross-platform of quantum transport implementation. The non-self-consistent step is indicated by the dark (blue on line) arrow.

This process of computing DM from H and vice versa iterates until the charge density converges, after which the S-matrix and transmission T(E) are computed for the converged density. Presently, we find that simple mixing of the charge density suffices for NEGF calculations, but more sophisticated mixing algorithms can easily be incorporated. We implemented the NEGF loop and mixing using a Python script, so each TiMeS and H determination step run as independent modules. The non-self-consistent and the NEGF flowchart are shown in Figure 2. This flexibility and modularity allows for easy continuation of interrupted runs, as well as interchange and comparison of different electronic structure approaches or density functionals. In the present work, matrix representations of the Hamiltonians of the relaxed structures were obtained from OpenMX as described in Section

2.1. The program flow is specifically designed to allow a modular incorporation of differing electronic structure representations of the single-particle Hamiltonian.

## 3. Results and discussion

### 3.1 *Transport properties*

We discuss the impact of numerical atomic orbitals in materials transport properties focusing on charge carrier scattering in SiNWs due to common impurities. The example of Ref.14 is taken as initial system of reference, where there is a detailed discussion on the scattering behavior due to oxygen defects varying the oxidation state of the Si surface. In this model the surface Si atom is locally oxidized to the formally $Si^{2+}$ state by forming a Si-O-Si back bond and with a hydroxyl instead of hydrogen for passivating the surface dangling bond (see inset of Figure 3). The presence of the oxygen defect is the origin for surface roughness at the atomic scale.

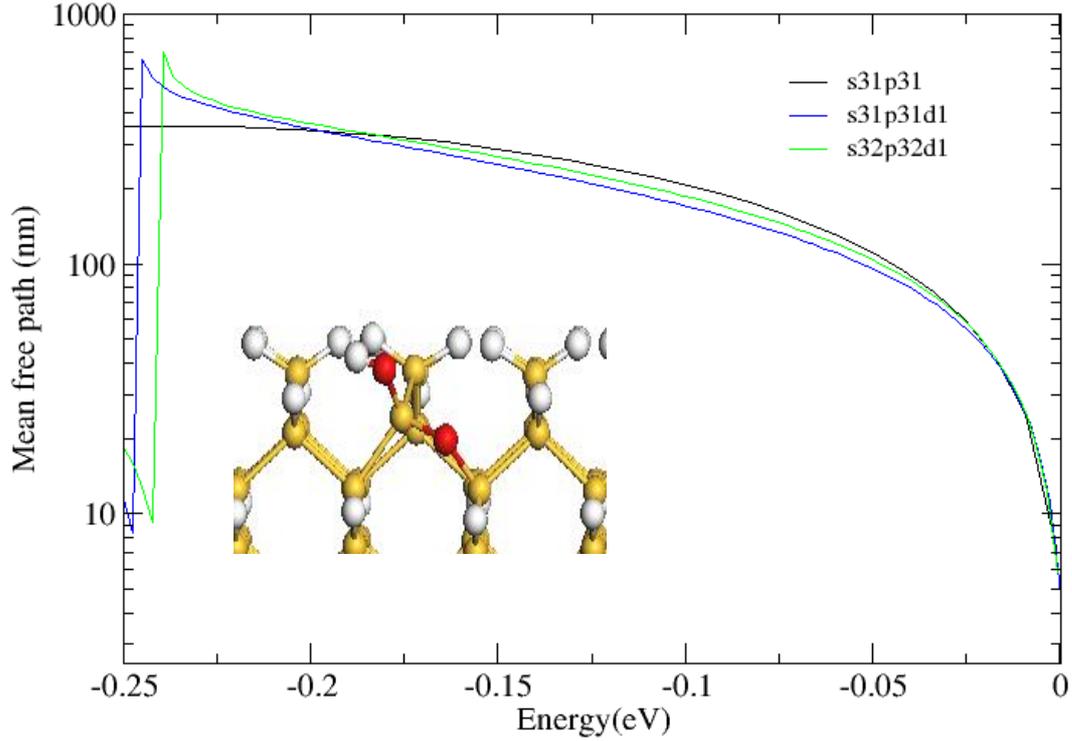

**Figure 3.** Mean free path of locally oxidized Si nanowire (structure described in text) for the indicated basis sets and defect density n = $5 \times 10^{19}$ cm$^{-3}$ (mean distance between impurities $l_d$ = 9.6 nm). The energy range corresponds to the first valence sub-band of the electronic structure obtained with the double-zeta polarized basis set.

Following the method in Ref. 46, the mean free path for electrons injected at energy E is estimated from $\lambda(E) = \frac{T_s(E)}{T_c(E)} \cdot l_d$ , where $T_s$ and $T_c$ are the transmission values across the wire with and without local oxidation, respectively and $l_d$ is the mean distance between defects. The mean free path in the energy range of the first valence sub-band is shown in Figure 1 for various basis sets, namely, single-zeta, single-zeta polarized and double-zeta polarized. Here, $l_d$ = 9.6 $nm$ which corresponds to a defect density n = $5 \times 10^{19}$ $cm^{-3}$. There is in overall good agreement with Ref. 14 where a minimal basis set was used within the Density Functional Tight Binding approximation. Interestingly, the mean free path obtained from the various basis sets shows variation of just up to 0.01% within an energy range of 0.2 eV using the double-zeta polarized results as a reference.

The above transport results confirm the weak dependence of λ on the basis set when the prevalent scattering mechanism is non-resonant scattering which is regularly observed for typical dopant impurities[46,47] and oxidation defects.[14] Using a simple analysis based on band structures, previous predictions attribute mean free path variations to small changes in the group velocity.[28] It may be expected that differing group velocity and effective mass will have the most significant impact in the transport coefficients when different electronic structures are used, hence, the need to calibrate to the experiment is introduced.

To consider the case of strong scattering, we study the common example of a boron substitutional impurity which can act as a p-type dopant.[47,48] The transmission of holes injected at energies within the first-valence sub-band is plotted in Figure 4. Calculations using single-zeta, single-zeta polarized, double-zeta polarized and triple-zeta polarized bases are shown. There is overall qualitative agreement between the various basis sets notwithstanding the evident resonant backscattering that strongly suppresses transmission. However, the quantitative discrepancy in the transmission may lead to significant overestimations of the nanowire conductivity. Similar results are found for an n-type dopant impurity, namely, substitutional phosphorus.

It is evident from Figure 4 that using the basis set with double-zeta and polarization closely tracks the result obtained with triple-zeta polarized basis. Hence the double-zeta basis may be viewed as a good compromise between size and accuracy in the application of atomic basis sets to predict from first-principles the characteristic transport length scales and intrinsic transport properties in materials. This behavior is similar to conductance estimations in transport across molecular junctions where convergence needs to be ensured by enlarging the size of the basis set. [30,31]

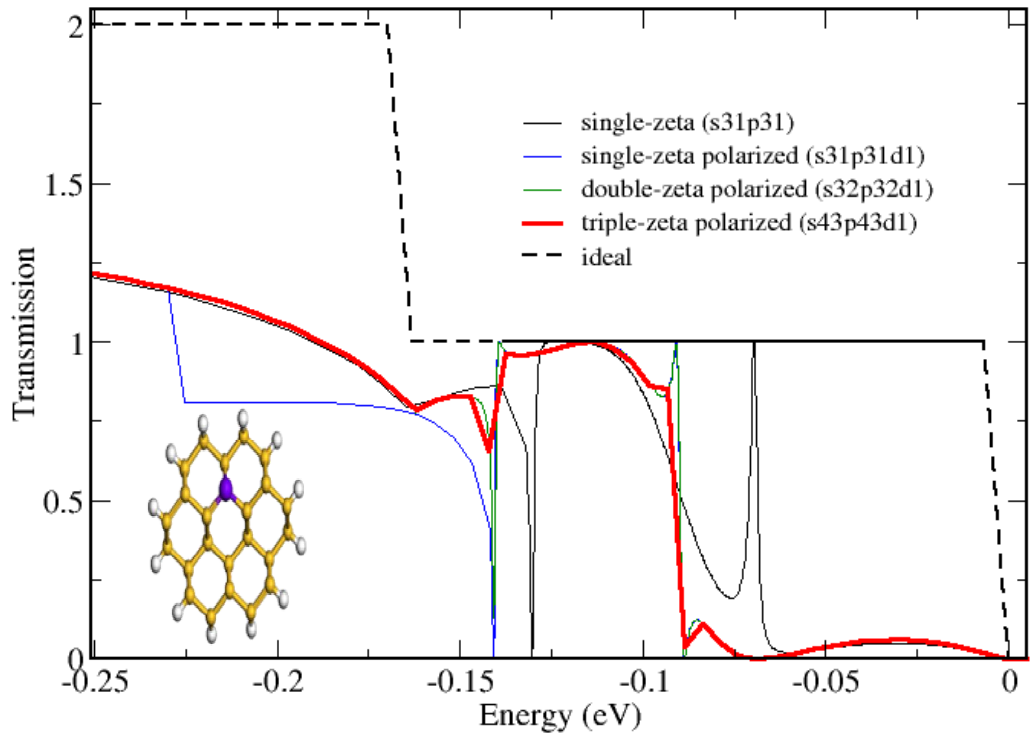

**Figure 4.** Transmission of holes across a silicon nanowire with a boron impurity for the indicated basis sets. The structure and the contraction scheme are described in the text. The energy range corresponds to the first valence sub-band of the electronic structure obtained with the double-zeta polarized basis set.

3.2 *Electrical characteristics*

Applying our electronic transmission methodology, in this section the electrical characteristics of SiNW based p-n-p and p-i-p junctions will be explored. The p-n-p junction is constructed from boron doped leads with an n-type scattering region between the two p-type leads (see Figure 1). The scattering region includes a single phosphorus dopant in the nanowire lattice. For the p-i-p junctions, the scattering region is intrinsic. The total length of the scattering region is 2.7 nm. In a nanowire-based FET the intrinsic region is surrounded by a cylindrical gate to yield a GAA configuration. Given that the localization radius of dopant impurity states is approximately 1.5 nm and that the channel length should be at least two times larger than the nanowire diameter[49], these models describe the smallest junctioned nanowire transistors that can be envisioned using conventional doping strategies. The results are obtained using the two different transport algorithms which have already been described in Section 2.2.

Figure 5 illustrates the transmission properties of the p-n-p and p-i-p SiNW junctions at different source-drain bias voltages, applied along the nanowire axis, as predicted from different orbital basis sets. As expected, Figure 5 shows that at small source-drain bias voltages the self-consistent converged transmission $T(E,V_{DS})$ does not differ from the linear-response approximation $T(E)$ (curve indicated as $V_{DS} = 0$) considerably. For larger bias applying the NEGF self-consistence loop increases the transmission for all basis sets used. The larger basis has an effect on the electronic structure alignment of the scattering region with the available channels of the leads as seen by the comparison of the transmission onset between minimal (Figures 4 (a) and (c)) and polarized basis sets (Figures 4 (b) and (d)). The transmission is also affected at energies well below the top of the valence band. These differences on the energy dependence of transmission result in variation in the current estimates as shown below. However, it is apparent that the addition of d-polarization functions does not significantly alter the self-consistence dependence of the transmission on applied voltage.

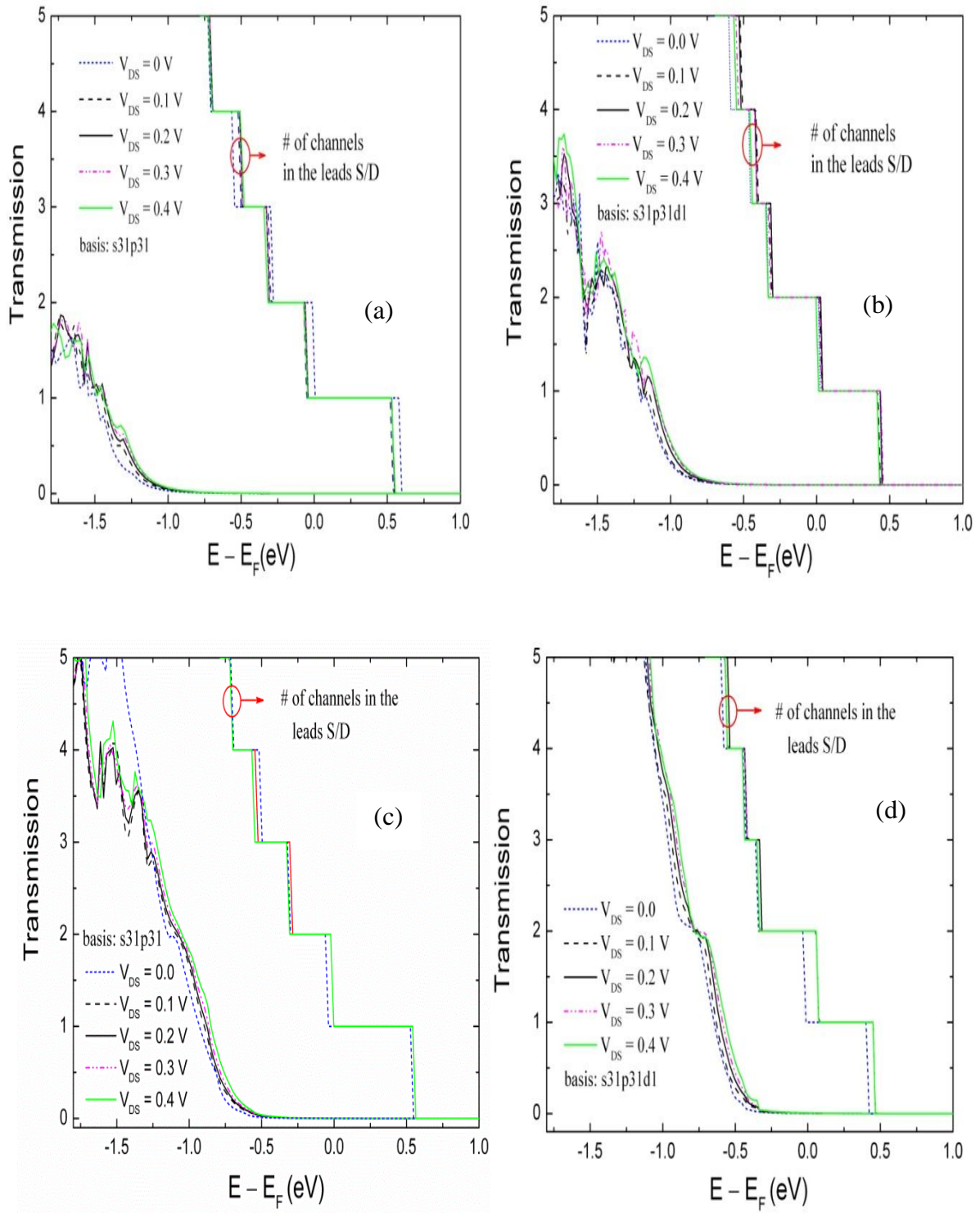

**Figure 5.** Transmission properties of p-n-p junctions using (a) s31p31 and (b) s31p31d1 basis sets. Panels (c) and (d) show the transmission of the p-i-p junctions considering s31p31 and s31p31d1 orbital basis sets, respectively. The Fermi level is the reference energy.

The current-voltage (I-V) characteristics of p-n-p and p-i-p junctions based on $T(E,V_{DS})$ and $T(E)$ are plotted in Figures 6 (a) and (b), respectively. The I-V characteristics using the double-zeta polarized basis set and without self-consistency (based on $T(E)$) is also illustrated in Figure 6 for comparison. The minimum orbital basis set (s31p31) predicts higher current flow for both junctions compared to orbital basis sets with polarization (s31p31d1). For example, at $V_{DS} = 0.4$ V, the p-n-p junction with s31p31 basis set, the current is 12% larger if we consider the current characteristic of s31p31d1 as the reference in the fully self-consistent approach. Also, this figure confirms that the difference between SC and non-SC results is greater at larger bias voltages for both structures as is intuitively expected.

The currents calculated with the non-SC loop are lower than applying the computationally more demanding NEGF method. This is attributed to an overestimation of the polarization effect induced when a bias is applied, whereas, SC allows for the redistribution of charges at non-equilibrium. This is important to take into account when simulating current-voltage characteristics to extract the device performance. Here, this yields an underestimation of the source-drain current. Elsewhere,[9] we have shown that estimates for switching between off- and on- states in nanowire transistors are much worse if self-consistency is disregarded in the calculation of the transfer characteristics. In that case there is no redistribution of the charges in the channel to lower the barrier as gate voltage is applied and the tunneling current is overestimated in the non-SC case. From Figure 5, it is deduced that the error is more significant than using a lower quality basis set.

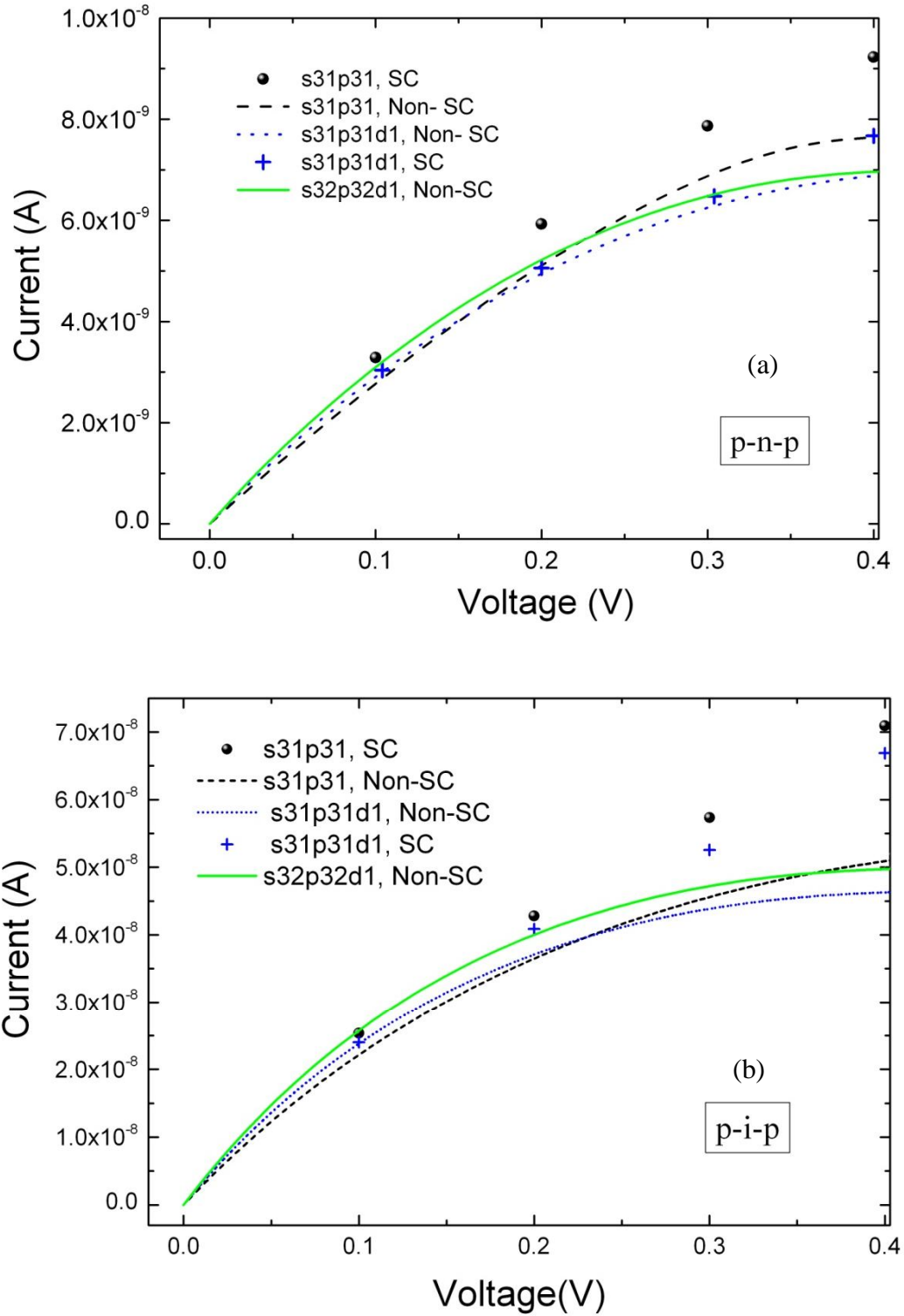

**Figure 6**. Current-voltage characteristics of (a) p-n-p and (b) p-i-p junctions. Different orbital basis sets are compared for both structures. The scattered points show results based on the self-consistent $T(E,V_{DS})$ and the lines correspond to the linear response approximation $T(E)$.

The I-V characteristic of the p-i-p junction shows more sensitivity at higher bias voltages compared to the p-n-p junction. The reason is that there is larger scattering for the p-n-p junction due to a larger in-built potential compared to the p-i-p junction (see Figure 5). This corresponds to higher level of injected electronic charges in the p-i-p junction yielding a higher current at large bias. Finally, the source-drain tunneling currents of the devices for the two doping profiles are around 0.5 nA and 2 nA for p-n-p and p-i-p junction, respectively. These values are two to four orders of magnitude higher compared to junctionless Si nanowire devices,[9,25] that is, FETs with homogeneous source-channel-drain doping similar channel dimensions and NW orientation. Supporting our previous findings[25] this suggests that at such scales the distribution of charge carriers around the dopant blurs the boundaries of a junction over a distance comparable to the channel length. Therefore, besides being very difficult to fabricate, junctioned FET designs at this scale will fail to effectively keep carriers out of the channel in the off-state.

### 4. Concluding remarks

To summarise, in this paper we evaluate the impact of numerical atomic orbitals on electron transport properties mainly on charge carrier scattering in [110] oriented SiNWs with dopant impurities. A relatively weak dependence of the mean free path on the NAO basis set was found using explicit calculations of quantum transport that combine DFT with the Green's function formalism. This is in agreement with estimates based on a simple band analysis. For the case of weak scattering from impurities the choice of the NAOs basis is considerably less important than for impurities inducing strong backscattering. It is found that optimised double-zeta orbitals with d polarisation functions reproduce the results with triple-zeta polarised basis; even in the case of suppressed transmission the double zeta polarised basis offers a good compromise between accuracy and computational efficiency. The single zeta basis set yields only qualitatively correct results.

We also assess the impact of various numerical atomic orbital approximations on the transport properties of silicon nanowire setups with p-n-p and p-i-p doping profiles. Interestingly, although the current-voltage characteristics of such ultrascaled (3 nm length) nanowire-based transistors show the expected sensitivity to the choice of the NAOs basis set, it is found that charge self-consistency affects mostly the source-drain current. Regarding the possibility to use these devices, it is concluded that the source-drain tunneling currents are relatively high – of the order of 0.5 nA and 2 nA for p-n-p junction and p-i-p junction, respectively) – making it difficult to achieve good device performance for these architectures on the length scale considered.

AUTHOR INFORMATION

**Corresponding Author**

*Giorgos Fagas, Georgios.Fagas@Tyndall.ie

**Author Contributions**

The manuscript was written through contributions of all authors. All authors have given approval to the final version of the manuscript. ‡These authors contributed equally.

**Funding Sources**

Science Foundation Ireland (SFI) under the Principal Investigator Grant No. 06/IN.1/I857.

European Union 7[th] Framework ICT-FET-Proactive program, SiNAPS project under contract No. 257856.

ACKNOWLEDGMENT

This research was funded by Science Foundation Ireland (SFI) under the Principal Investigator Grant No. 06/IN.1/I857. Partial support was provided through the European Union 7th Framework ICT-FET-Proactive program, SiNAPS project under contract No. 257856. We also acknowledge computing resources provided by SFI to the Tyndall National Institute and by the SFI and Higher Education Authority Funded Irish Centre for High End Computing.